# MUSIC GENERATION WITH TEMPORAL STRUCTURE INFORMATION

**Shakeel ur Rehman Raja**
shakeelraja@city.ac.uk

**ABSTRACT**

In this paper we introduce a novel feature augmentation approach for generating structured musical compositions comprising melodies and harmonies. The proposed method augments a connectionist generation model with *count-down to song conclusion* and *meter markers* as extra input features to study whether neural networks can learn to produce more aesthetically pleasing and structured musical output as a consequence of augmenting the input data with structural features. An RNN architecture with LSTM cells is trained on the Nottingham folk music dataset in a supervised sequence learning setup, following a Music Language Modelling approach, and then applied to generation of harmonies and melodies. Our experiments show an improved prediction performance for both types of annotation. The generated music was also subjectively evaluated using an on-line Turing style listening test which confirms a substantial improvement in the aesthetic quality and in the perceived structure of the music generated using the temporal structure.

## 1. INTRODUCTION

Recent developments in the field of neural networks have allowed researchers and practitioners from many different areas of study to improve the state of the art for pattern recognition and machine learning (ML) tasks [17]. This development has also generated interest in the field of generative arts, where it allows provides computational models that can recognise complex patterns in data structures including spoken languages, visual objects and music. This intersection of art, creativity and technology has shown potential to understand and emulate human-like cognitive abilities. Within generative arts, music generation is considered to be particularly challenging problem domain as musical creativity involves complex multi-dimensional cognitive processes involving pitch, rhythm, harmony and higher-level structural inference and resulting subjective interpretation. These processes are not fully understood and are therefore very difficult to model in a computational framework [3]. A number of experiments been performed recently with deep generative models to overcome these challenges in music generation with varying level of success [5][6][9]. These experiments have shown that learning and demonstrating human-like understanding of musical structure towards composition is a highly challenging task and a rich area of research.

In this paper we improve the performance of a Recurrent Neural Network with Long Short Term Memory (LSTM) cells in a Language Modelling framework by augmenting the music data with temporal structure information. We add information about the beat positions and a count-down towards the song conclusion in training, prediction and generation. We apply this method in the prediction and stochastic generation of musical sequences containing melodies and harmonies and evaluate the objective performance in melody and harmony in prediction as well as th subjectively perceived structure and aesthetic quality in a human listening experiment. Both prediction and generation benefit from the augmented data, but the results show that prediction results are not sufficient to predict perceived musical quality and that more work on understanding musical structure and is needed to improve the quality of computer generated music.

Our main contributions in this paper are the development of an effective method for augmenting symbolic music representation with temporal structure information and the experimental results on prediction performance and the human evaluation of the music generated with our method.

The remainder of this article is structured as follows: Section 2 discusses related work from the literature, section 3 introduces our method, section 4 describes the experiments and discusses the results, section 5 presents the conclusion and future work.

## 2. RELATED WORK

During recent years, a number of experiments using Deep FFNs, RBMs, CNNs and Autoencoders have shown limitations of these architectures towards learning coherent musical structure due to difficulties in processing temporal structures [4]. Hybrid deep neural architectures have shown promising results including an RNN-RBM hybrid architecture [2], Generative Adversarial Networks (GANs) for symbolic music learning and generation [22] and Reinforcement Learning (RL) based RL-TUNER [9]. However, the level of complexity and computational cost required by these architectures pose limitations on their practical application, flexibility and scalability.

Recurrent Neural Networks (RNNs) have gained popularity in the field of music generation due to their ability retain a memory of processed temporal data. Mozer conducted the first major experiment for music generation using RNNs with CONCERT system [14]. The generated music was described as "*occasionally pleasant*" as it lacked a coherent structure due to the problem of exploding gradients as mentioned in [15]. Simple sigmoid based

RNNs seem to lack the ability to model structure of a composition. More sophisticated models like Long Short-Term Memory Networks (LSTMs) cells add complexity to a simple RNN cell by using gates to allow the cell to retain information for longer periods of time [8]. In 2002, Eck and Schmidhuber updated Mozer's approach by switching from a simple sigmoid activation based RNN to more sophisticated LSTM architecture as described in [7].

The true potential of RNNs towards generative arts, particularly in language modelling was highlighted by Andrej Karpathy [11], who showed that a simple RNN architecture called Char-RNN was able to recreate the "look and feel" of any input text corpus as sequence learning problem. Karpathy's approach towards text generation was applied to music by Bob Sturm to use Char-RNN with Irish folk music in ABC format with MLM [19]. Johnson modified the RNN architecture and created biaxial RNNs that used a note axis and a time axis to learn the temporal and structural dimensions [10]. Walder used LSTM networks with a custom encoding technique in an attempt to capture global structure [21]. In these experiments, LSTMs managed to capture some notion of local structure but the output still lacked a global structure and/or was predictable due heavy constraints.

Different encoding techniques and architecture are currently being employed by researchers and practitioners to enhance the learning process. Augmentation of domain specific knowledge into the training process is one approach which is heavily being applied to deep learning [12]. Feature learning and engineering approaches as mentioned by [1] are also gaining momentum with deep learning architectures. In music generation, this approach has been tried with the lookback RNN architecture in Google Magenta [20] that uses an attention mechanism when predicting next musical event.

## 3. APPROACH AND METHODS

This section will briefly discuss data representation, feature augmentation, network architecture, and music generation, with the methods chosen to fulfil stated research objective.

### 3.1 Baseline Model

We set up a baseline model first to compare the performance improvement in the following augmentation related experiments. The baseline model is inspired by an experiment by Yoav Zimmerman [23] that uses a "dual classification approach" to music language modelling as shown in Fig 1. This experiment uses the Nottingham Music dataset (Available at: http://www.chezfred.org.uk/freds/music ) to train an RNN architecture with a dual softmax loss function, taking into account melody and harmony loss simultaneously.

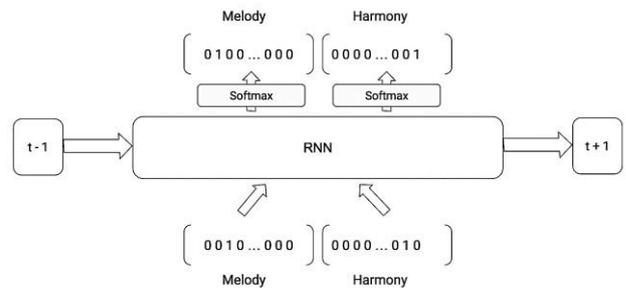

**Figure 1.** The dual softmax classification model [23].

*3.1.1 Language Modelling*

The baseline model operates in a *Music Language Model* (MLM) framework. A *Language Model* is a probability distribution over sequences of words:

$$P(w_1, .., w_m) = \prod_{i=1}^{m} P(w_i | w_1, .., w_{i-1}) \quad (1)$$

The model describes the likelihood of a word $w_i$ from a finite vocabulary $m$, given all previous words $w_1, ..., w_{i-1}$. For MLM, this approach has been used to model a probability distribution of musical notes as shown in [18]. Nottingham dataset is parsed with a monophonic assumption for melody i.e. only one note playing at a given time step. We also assume that harmony notes at each time step can be classified into as chord class which belongs to a finite chord dictionary containing major and minor chords for each note i.e. maximum 24 classes.

*3.1.2 Data representation*

The pre-processing stage maps a melody note as a discretized 35-bit one-hot vector with a pre-defined melody range (G3 – E6) at each time step. The parsing is performed with a predefined time step of 1/16 of a bar (1/4th of quarter beat in 4/4). In order to convert harmonic notes into their respective chords, Python Mingus library is used to determine the chord being played, before creating a second 23-bit one hot harmony vector (22 chords found in corpus, plus 1 class for unknown chords). The MIDI to piano roll mappings are created by appending melody and harmony one hot vectors on each time step and thus creating a final input encoding as shown in Fig. 2 (bit 0 – 57). Zero padding is applied to the corpus, based on the selected mini-batch length of 128 to preserve the length of the variable length sequences before mini-batching the dataset.

*3.1.3 Network Architecture*

We choose a simple LSTM architecture for training the network with Cross Entropy Loss to measure error between the network output and predicted values. RMSProp adaptive learning method is used for backpropagation learning. For final loss, we use a weighted sum of melody and harmony loss as suggested in [23]:

$$L(z, m, h) = \alpha \log\left(\frac{e^{z_m}}{\sum_{n=0}^{M-1} e^{z_n}}\right) + (1-\alpha)\log\left(\frac{e^{z_{M+h}}}{\sum_{n=M}^{M+H} e^{z_n}}\right) \quad (2)$$

where *M* and *H* are melody and harmony classes. The function calculates the log loss at a time step for the output layer $z \in R^{M+H}$, a melody label class *m*, and target harmony label class *h*, *α* termed as melody coefficient shifts the focus of loss function between memory and harmony. We set *α* to 0.5, giving equal importance to melody and harmony loss. **3.2 Data Augmentation**

We calculate additional features to augment the input encoding of the baseline model in order to provide extra information about the *duration, metre* and *rhythm* of MIDI sequences. These features are not used in the output and are not predicted or generated. In the generation phase, this set of features is also calculated and introduced into the generative process in what we call the *feature generator* (see Fig. 3).

*3.2.1 Song Conclusion Count-down*

The Song Conclusion Count-down is a counter based on song duration, measured in number of time steps, and appended to input encoding. This counter provides an indication of temporal structure by counting down towards song conclusion. The motivation is to help the network learn how melodies and harmonies change towards the end of the piece, or possibly even higher-level features, that as the counter approaches zero, melodies should be are similar to those played at the beginning. The counter values are normalized, as this led to faster learning in preliminary experiments.

The counter is rolled back so that zero points to the beginning of last note rather than the last time step in the MIDI sequences. This follows the musical intuition that the beginning of the last note is structurally more relevant than its ending point or its duration. Both variants both variants (zero at end of sequence or onset of last note) were tried in prelimiary tests and not noticable difference was found in performance. We therefore kept the last-noteonset variant, as it corresponds better to musical intuition.

*3.2.2 Augmenting Metrical Information*

In music, the meter, associated with recurring patterns of note accents and pitches is known to have a strong impact on listeners' perception well as future anticipation of musical patterns "that we abstract from the rhythm surface of the music as it unfolds in time" [13]. Following this motivation, we augment metrical information to the input encoding with meter-markers as follows. We create two onehot vectors, first to count each time step towards a quarter beat (4 time steps or sixteenth notes), and a second counter to count quarter beats up to whole bar (4 quarter beats for a 4/4 time signature). We also test for the presence of an anacrusis by taking the total length modulo the bar length, which turned out to be a good heuristic on this dataset, and started appending metrical features from the first downbeat as shown in Fig 2.

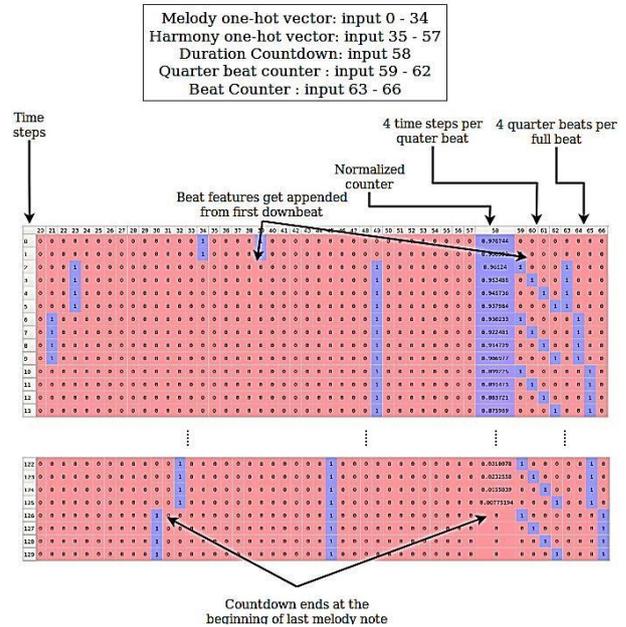

**Figure 2.** Input encoding with engineered duration and metrical feature augmentation.

**3.3 Generation**

For generation, we use conditioning and sampling techniques suggested by [4] and [7], in order to sample from the probability distributions generated by our model. The MLM approach is used to define the probability distribution over a sequence *x* by using the predictions from previous time step *t-1* as inputs to the next time step *t*. The output at time *t* is used as inputs at time *t+1* in a similar fashion to create a generation loop (see Fig 3). For generation, we set a sequence length of 384 timesteps (24 4/4 bars). A feature generator, as shown in Fig 3, is programmed into the generation stage to provide augmented structural features like those in the training dataset. Based on the defined length of generated sequences, count-down and meter-markers are calculated and appended to the input sequences during conditioning and generation phases using a 4/4 metre. For comparison, we use the standard MLM generation approach without the feature generator.

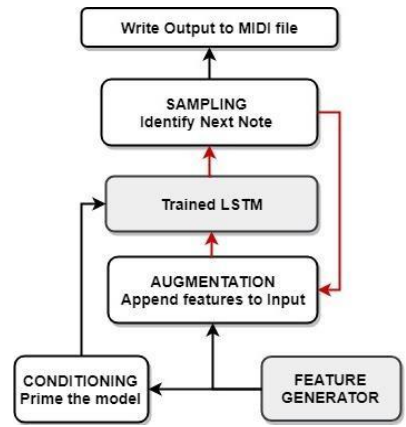

**Figure 3.** The generation framework. A feature generator is introduced in the generative process to calculate and append engineered features for conditioning and sampling from the output.

## 4. EXPERIMENTAL PROCEDURE AND RESULTS

### 4.1 Objective Evaluation – Prediction Performance

We conduct a number of experiments using the framework described earlier. With the Nottingham dataset containing almost 1000 sequences, we created a 70 – 30 train and test split with stratified sampling, ensuring a balanced distribution of included folk styles. For each experiment, a grid search is performed to identify the best hyper-parameters for the model. An early stopping criterion is applied that stops the training process if no improvement in validation loss is seen for 15 epochs. The following ranges are used in the grid search:

**Number of Hidden Layers:** [1, 2] **Hidden Layer Size:** [100, 150, 200] **Drop Out Keep Probability:** [0.75, 0.5, 0.3]

Table 1 shows a comparison of loss performances of best models obtained from parameter grid searches in augmentation experiments performed with the setup described earlier.

|     | Layers | Hid Size | Drop Out | Train Loss | Valid Loss | Time/Epoch |
|-----|--------|----------|----------|------------|------------|------------|
| BL  | 2      | 150      | 0.5      | 0.29601    | 0.51209    | 74.48      |
| CD  | 2      | 100      | 0.5      | 0.39425    | 0.47357    | 58.57      |
| MM  | 2      | 200      | 0.3      | 0.30572    | 0.48993    | 82.54      |
| FC  | 2      | 200      | 0.3      | 0.34760    | 0.46932    | 116.5      |
| 4/4 | 2      | 200      | 0.5      | 0.27501    | 0.46910    | 85.17      |

**Table 1. BL** = Baseline Model, **CD** = Count-down augmented model, **MM** = Meter-markers augmented model, **FC** = Combined Augmentation (CD + MM, full dataset), **4/4** = Combined Augmentation (CD + MM, 4/4 time signature only)

#### 4.1.1 Baseline Model (BL)

We train and test the dual SoftMax LSTM architecture without any feature augmentation as the baseline model. The best model performance on the augmented dataset shows a training loss of 0.2960 and a validation loss of 0.5121. Our result is slightly lower than that reported in the original experiment [23], where the best validation loss is 0.536. We suspect that this difference this is an effect of zero-padding we used to equalise the song lengths whilte the original experiment truncated the songs for this purpose. A dropout of 0.5 and early stopping prevents the network from overfitting and training converges much faster than original experiment (40 epochs vs 250 epochs in original experiment [23]). We use this model as the baseline for the remaining experiments.

#### 4.1.2 Song Conclusion Count-down Augmentation (CD)

We append the count-down feature to the input encoding of the baseline model from the previous experiment. The parameter grid remains the same thoughout the experiments. This augmentation experiment shows a reduced validation loss of 0.4736. Interestingly, a higher training error is observed with count-down augmentation, along with a reduced training time. This may be an indication that the model learns additional information due to augmented count-down feature which prevents the model from overfitting.

#### 4.1.3 Meter-marker Augmentation (MM)

In order to monitor the effect of meter-markers on the LSTM's predictive performance, we append metrical features to the input encoding of the baseline model. Initially, using two one hot vectors for the quarter note and the sixteenth note level (quarter beats, see 3.2.2), a poor model performance is observed along with poor aesthetic quality of generated sequence. Considering that quarter beat markers maybe conflicting with LSTM's ability to learn note transition probabilities defining short term structure [7], we decide to remove quarter beat markers from input encoding and train the model with a whole beat marker only. The best model from the grid search shows a training loss of 0.3057 and validation loss of 0.4899. The meter marker is always in a 4/4 structure, we have run a second experiment where we use only the music in 4/4 times signature, as explained in section 4.1.5

#### 4.1.4 Count-down and Meter-marker Augmentation (CM)

Next, we apply both augmentations from 4.1.2 and 4.1.3 simultaneously to observe the combined effect of these annotations. The best model shows a lower validation error of 0.4693 with training error of 0.3476. As observed in the last experiment (MM), this performance is obtained with a dropout of 0.3, showing a need to substantial regularisation.

#### 4.1.5 Feature Augmentation with 4/4 Music (4/4)

We started with metre markers in 4/4 time signature as a first approximation, since the majority of pieces in our dataset has an time signature with a power of 2 in the numerator (4/4,2/4,2/2). We noticed that the prediction performance did not differ much between the pieces in 4/4 time signature and the other pieces, where there is a factor of 3 in the numerator (3/2,3/4,6/4,6/8,9/8). However, in the generation phase we noticed that there were clearly perceivable differences. We therefore evaluated the 4/4 pieces separately where the metre marker matches the time signature.

We select sequences with 4/4 time signature only, and try the count-down metrical augmentation which exactly matches the metre of these sequences. The best model from the grid search shows a training loss of 0.2750 and validation loss 0.4691. The reduced training loss is probably due to reduced data size as 4/4 sequences roughly make up about 50 percent of Nottingham dataset, leading to some overfitting. Due to this reduction, the model cannot be directly compared to previous models' performance but generated sequences will be used to evaluate this model further.

Overall the observations from the objective evaluation show that augmented features have a positive impact as both annotations yield an reduction of the loss.

### 4.2 Subjective Evaluation

We use a number of music sequences generated with the framework described earlier for subjective evaluation through an online Turing style listening test. We use five linear scale type questions with a range of 1 to 5 to measure the level of improvement for each sequence. The questions address:

1. Likelihood of sequence being composed by a machine.
2. Quality of long term structure.
3. Quality of short term structure
4. Quality of song conclusion
5. Overall aesthetic quality

On the linear scale, 1 represents *human/pleasant* and 5 represents computer *generated/poor*. Some elements for developing the listening test are taken from the framework developed by Pearce and Wiggins [16]. As the output of the network is stochastic, the generated sequences vary in quality. To address both a scenario of human selection and one of unsupervised music generation, one randomly selected and one hand-picked sequence are used in the test, for all intermediate and final experiments, in addition to two randomly chosen human compositions from the Nottingham dataset. We present these sequences to a total of 40 listeners in a different presentation orders during the test. User groups are identified as; professional musicians and enthusiasts (Group 1: 12.5%), understand music theory or play a musical instrument (Group 2: 25%), casual listeners (Group 3: 60%) and listeners with no interest in music (Group 4: 2.5%).

Fig. 4 shows the average scores for these sequences given by each listener group. Averaged listener scores show that listeners from groups 1 and 2 can easily differentiate between computer generated music and human compositions, as opposed to groups 3 and 4. It is evident that sequences with count-down augmentation get better overall results (lower values) when compared to the base model. Meter-marker augmentation did not show any substantial improvement when augmented with complete dataset. An overall improvement, can be seen when structural features are appended to 4/4 music only.

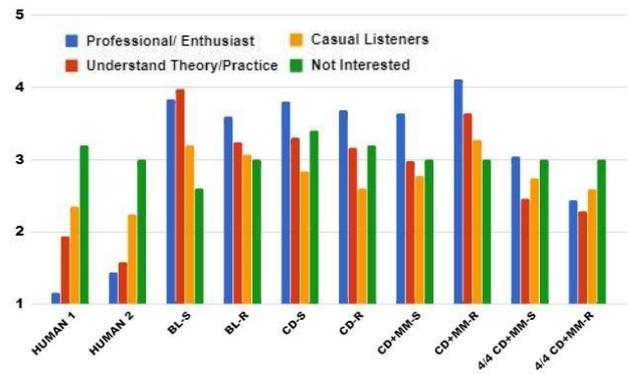

**Fig 4. Human 1 & 2** = Human Compositions, **BL** = Baseline Model, **CD** = Count-down Augmentation, **MM** = Meter-markers Augmentation. **-R** = Random Sequences, **-S** = Selected Sequence.

For further investigation, we use randomly generated sequences from the baseline model, count-down augmentation, count-down + meter-marker augmentation (4/4) along with one human composition for comparison. In order to study the impact of augmented features on song structure and resulting quality, we only consider feedback from listeners having musical knowledge (Group 1 and 2).

*4.2.1 Baseline Model*

Fig 5a shows user scores for the randomly selected sequence generated from the baseline model. This model is not well rated, and rather clearly identified as computer generated music in terms of aesthetic quality. The sequence shows poor score for musical structure, conclusion and overall quality. This concurs also with our observation that the sequences produced by this model end abruptly, making them sound even less pleasant.

*4.2.2 Human Composition*

Fig. 5b shows that listeners from group 1 and 2 gave a high score to the human compositions identifying them with good aesthetic quality, a well-defined long and short-term structure and a pleasant conclusion. A slightly low score in long term structure and overall quality of the sequence is thought to be a result of listeners' personal preferences towards folk music.

*4.2.3 Count-down Augmentation*

With the introduction of count-down, the overall rating is more positive compared to the baseline model. Listeners higher than baseline model as seen in Fig 5c. The countdown augmentation appears to allow the model to learn more temporal structure information and generate sequences with a somewhat improved structure towards song conclusion. The augmentation has no effect on short term structure.

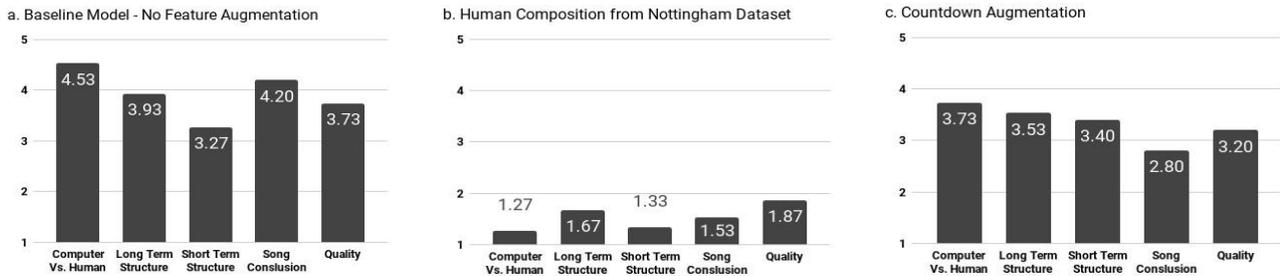

**Figure 5.** Subjective evaluation of baseline model, human composition and countdown augmentation. Countdown augmentation shows an improved performance towards song conclusion and overall aesthetic quality.

s

*4.2.4 Count-down and Meter Markers Augmentation*

As already described in section 4.1.5, applying both features only to 4/4 music leads to generation of sequences which are structurally more sound with an improved aesthetic appeal. The sequences are rated as having more human-like qualities showing an better score than for music generated by the model trained on the full dataset for most areas as seen in Fig 6**.** In our observation, the generative model maintains the meter for most parts, and occasionally repeats rhythmic patterns for the sequence duration as a result of meter-marker augmentation.

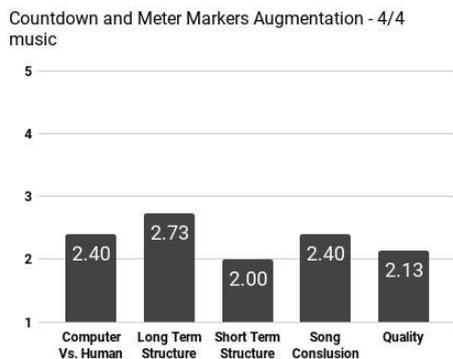

**Figure 6.** Subjective evaluation of countdown and meter marker-based composition by listeners from Group 3, 4.

The model learns to introduce more suitable short-term note transitions giving the sequence a stronger sense of a coherent rhythm, which associates it more with human compositions. The songs conclude well, we believe due to count-down augmentation.

## 5. CONCLUSION AND FUTURE WORK

Our results show that augmenting temporal structural information to input encoding positively affects the performance of the model. The *count-down to song conclusion* counter augmentation shows a clear improvement in the predictive performance of the model with lower loss values as shown by results. Subjective evaluation of model output also shows a higher score for song conclusion and overall quality when compared to the baseline model. *Meter-marker* augmentation only contributes substantially towards loss improvement when the sequences of matching time signature exactly (4/4 in this case), although the difference in prediction performance is relatively small. An improved score in final experiment is shown, especially in terms of short term structure and an improved song conclusion. We conclude also that prediction loss of a music language model may not directly relate to the aesthetic quality of generated sequences, given the discrepancy between the huge improvement in subjective evaluation and the small improvement in prediction caused by using only 4/4 music with the model. Our results confirm the effectiveness of chosen approaches and methods towards improving short term structure and duration control and highlight further research potential of this approach with architectures described in [5],[6] and [9].

We plan to scale up this experiment by increasing the size of training data. A more flexible pre-processing routine will be formulated to in order to deal with more types of data so that different datasets can be employed. A more rigorous testing pipeline with cross-validation will also be implemented, and further listening tests and in-depth analyses of generated music are planned. The chosen approach will also be tried with other more sophisticated RNN architectures, such as Generative Adversarial Networks [24].

Python code and audio samples available at: https://goo.gl/boRVvx